
\input epsf
\catcode`\@11\relax
\newif\ify@autoscale \y@autoscaletrue \def\Yautoscale#1{\ifnum #1=0
  \y@autoscalefalse\else\y@autoscaletrue\fi}
\newdimen\y@b@xdim
\newdimen\y@boxdim \y@boxdim=13pt
\def\Yboxdim#1{\y@autoscalefalse\y@boxdim=#1}
\newdimen\y@linethick    \y@linethick=.3pt
\def\Ylinethick#1{\y@linethick=#1}
\newskip\y@interspace \y@interspace=0ex plus 0.3ex
\def\Yinterspace#1{\y@interspace=#1}
\newif\ify@vcenter   \y@vcenterfalse
\def\Yvcentermath#1{\ifnum #1=0 \y@vcenterfalse\else\y@vcentertrue\fi}
\newif\ify@stdtext   \y@stdtextfalse
\def\Ystdtext#1{\ifnum #1=0 \y@stdtextfalse\else\y@stdtexttrue\fi}
\newif\ify@enable@skew   \y@enable@skewfalse
\expandafter\ifx\csname enableskew\endcsname\relax
 \y@enable@skewfalse \else \y@enable@skewtrue\fi
\def\y@vr{\vrule height0.8\y@b@xdim width\y@linethick depth 0.2\y@b@xdim}
\def\y@emptybox{\y@vr\hbox to \y@b@xdim{\hfil}}
\ify@enable@skew
 \def\y@abcbox#1{\if :#1\else
   \y@vr\hbox to \y@b@xdim{\hfil#1\hfil}\fi}
 \def\y@mathabcbox#1{\if :#1\else
   \y@vr\hbox to \y@b@xdim{\hfil$#1$\hfil}\fi}
\else
 \def\y@abcbox#1{\y@vr\hbox to \y@b@xdim{\hfil#1\hfil}}
 \def\y@mathabcbox#1{\y@vr\hbox to \y@b@xdim{\hfil$#1$\hfil}}
\fi
\def\y@setdim{%
  \ify@autoscale%
   \ifvoid1\else\typeout{Package youngtab: box1 not free! Expect an
     error!}\fi%
   \setbox1=\hbox{A}\y@b@xdim=1.6\ht1 \setbox1=\hbox{}\box1%
  \else\y@b@xdim=\y@boxdim \advance\y@b@xdim by -2\y@linethick
  \fi}
\newcount\y@counter
\newif\ify@islastarg
\def\y@lastargtest#1,#2 {\if\space #2 \y@islastargtrue
  \else\y@islastargfalse\fi}
\def\y@emptyboxes#1{\y@counter=#1\loop\ifnum\y@counter>0
  \advance\y@counter by -1 \y@emptybox\repeat}
\def\y@nelineemptyboxes#1{%
  \vbox{%
    \hrule height\y@linethick%
    \hbox{\y@emptyboxes{#1}\y@vr}
    \hrule height\y@linethick}\vskip-\y@linethick}
\def\yng(#1){%
  \y@setdim%
  \hskip\y@interspace%
  \ifmmode\ify@vcenter\vcenter\fi\fi{%
  \y@lastargtest#1,
  \vbox{\offinterlineskip
    \ify@islastarg
     \y@nelineemptyboxes{#1}
    \else
     \y@ungempty(#1)
    \fi}}\hskip\y@interspace}
\def\y@ungempty(#1,#2){%
  \y@nelineemptyboxes{#1}
  \y@lastargtest#2,
  \ify@islastarg
   \y@nelineemptyboxes{#2}
  \else
   \y@ungempty(#2)
  \fi}
\def\y@nelettertest#1#2. {\if\space #2 \y@islastargtrue
  \else\y@islastargfalse\fi}
\def\y@abcboxes#1#2.{%
  \ify@stdtext\y@abcbox#1\else\y@mathabcbox#1\fi%
  \y@nelettertest #2.
  \ify@islastarg\unskip%
   \ify@stdtext\y@abcbox{#2}\else\y@mathabcbox{#2}\fi%
  \else\y@abcboxes#2.\fi}
 \newdimen\y@full@b@xdim
 \newcount\y@m@veright@cnt
\ify@enable@skew
 \def\y@get@m@veright@cnt#1#2.{%
   \if :#1 \advance\y@m@veright@cnt by 1\y@get@m@veright@cnt#2.\fi}
 \let\y@setdim@=\y@setdim
 \def\y@setdim{%
   \y@setdim@ \y@full@b@xdim=\y@b@xdim
   \advance\y@full@b@xdim by 1\y@linethick}
 \def\y@m@veright@ifskew#1{
   \y@m@veright@cnt=0 \y@get@m@veright@cnt#1.
   \moveright \y@m@veright@cnt\y@full@b@xdim}
\else
 \def\y@m@veright@ifskew#1{}
\fi
\def\y@nelineabcboxes#1{%
  \y@nelettertest #1.
  \ify@islastarg
   \y@m@veright@ifskew{#1}
    \vbox{
      \hrule height\y@linethick%
      \hbox{\ify@stdtext\y@abcbox#1\else\y@mathabcbox#1\fi\y@vr}
      \hrule height\y@linethick}\vskip-\y@linethick
  \else
   \y@m@veright@ifskew{#1}
    \vbox{
      \hrule height\y@linethick%
      \hbox{\y@abcboxes #1.\y@vr}%
      \hrule height\y@linethick}\vskip-\y@linethick
  \fi}
\def\young(#1){%
  \y@setdim%
  \hskip\y@interspace%
  \y@lastargtest#1,
  \ifmmode\ify@vcenter\vcenter\fi\fi{%
  \vbox{\offinterlineskip
    \ify@islastarg\y@nelineabcboxes{#1}%
    \else\y@ungabc(#1)%
    \fi}}\hskip\y@interspace}
\def\y@ungabc(#1,#2){%
  \y@nelineabcboxes{#1}%
  \y@lastargtest#2,
  \ify@islastarg\y@nelineabcboxes{#2}%
  \else\y@ungabc(#2)%
  \fi}
\catcode`\@12\relax
 

\newfam\scrfam
\batchmode\font\tenscr=rsfs10 \errorstopmode
\ifx\tenscr\nullfont
        \message{rsfs script font not available. Replacing with calligraphic.}
        \def\scr{\cal}
\else   
        \font\sevenscr=rsfs7
        \font\fivescr=rsfs5
        \skewchar\tenscr='177 \skewchar\sevenscr='177 \skewchar\fivescr='177
        \textfont\scrfam=\tenscr \scriptfont\scrfam=\sevenscr
        \scriptscriptfont\scrfam=\fivescr
        \def\scr{\fam\scrfam}
        \def\cal{\scr}
\fi
\catcode`\@=11
\newfam\frakfam
\batchmode\font\tenfrak=eufm10 \errorstopmode
\ifx\tenfrak\nullfont
        \message{eufm font not available. Replacing with italic.}
        
\else
	
	\font\sevenfrak=eufm7 \font\fivefrak=eufm5
	\textfont\frakfam=\tenfrak
	\scriptfont\frakfam=\sevenfrak \scriptscriptfont\frakfam=\fivefrak
	
\fi
\catcode`\@=\active
\newfam\msbfam
\batchmode\font\twelvemsb=msbm10 scaled\magstep1 \errorstopmode
\ifx\twelvemsb\nullfont\def\Bbb{\bf}

	\message{Blackboard bold not available. Replacing with boldface.}
\else   \catcode`\@=11
        \font\tenmsb=msbm10 \font\sevenmsb=msbm7 \font\fivemsb=msbm5
        \textfont\msbfam=\tenmsb
        \scriptfont\msbfam=\sevenmsb \scriptscriptfont\msbfam=\fivemsb
        \def\Bbb{\relax\expandafter\Bbb@}
        \def\Bbb@#1{{\Bbb@@{#1}}}
        \def\Bbb@@#1{\fam\msbfam\relax#1}
        \catcode`\@=\active

\fi
        \font\eightrm=cmr8              \def\xrm{\eightrm}
        \font\eightbf=cmbx8             \def\xbf{\eightbf}
        \font\eightit=cmti10 at 8pt     \def\xit{\eightit}
                     
        \font\eightcp=cmcsc8
        \font\eighti=cmmi8              \def\xold{\eighti}
        \font\eightib=cmmib8             \def\xbold{\eightib}
        \font\teni=cmmi10               \def\old{\teni}
        \font\tencp=cmcsc10

        \font\twelvecp=cmcsc10 scaled\magstep1

        \font\eightmath=cmmi8

	 at10pt	
	\font\twelvehelvbold=phvb at12pt
	 at14pt
	\font\sixteenhelvbold=phvb at16pt

\def\noblackbox{\overfullrule=0pt}
\noblackbox

\newtoks\headtext
\headline={\ifnum\pageno=1\hfill\else
	\ifodd\pageno{\eightcp\the\headtext}{ }\dotfill{ }{\old\folio}
	\else{\old\folio}{ }\dotfill{ }{\eightcp\the\headtext}\fi
	\fi}
\def\makeheadline{\vbox to 0pt{\vss\noindent\the\headline\break
\hbox to\hsize{\hfill}}
        \vskip2\baselineskip}
\newcount\infootnote
\infootnote=0
\def\foot#1#2{\infootnote=1
\footnote{${}^{#1}$}{\vtop{\baselineskip=.75\baselineskip
\advance\hsize by -\parindent\noindent{\xrm #2}}}\infootnote=0$\,$}
\newcount\refcount
\refcount=1
\newwrite\refwrite
\def\oldsize{\ifnum\infootnote=1\xold\else\old\fi}
\def\ref#1#2{
	\def#1{{{\oldsize\the\refcount}}\ifnum\the\refcount=1\immediate\openout\refwrite=\jobname.refs\fi\immediate\write\refwrite{\item{[{\xold\the\refcount}]} 
	#2\hfill\par\vskip-2pt}\xdef#1{{\noexpand\oldsize\the\refcount}}\global\advance\refcount by 1}
	}
\def\refout{\catcode`\@=11
        \xrm\immediate\closeout\refwrite
        \vskip2\baselineskip
        {\noindent\twelvecp References}\hfill\vskip\baselineskip
        \baselineskip=.75\baselineskip
        \input\jobname.refs
        \baselineskip=4\baselineskip \divide\baselineskip by 3
        \catcode`\@=\active\rm}

\def\hepth#1{\href{http://arxiv.org/abs/hep-th/#1}{hep-th/{\xold#1}}}
\def\mathdg#1{\href{http://arxiv.org/abs/math.DG/#1}{math.DG/{\xold#1}}}
\def\jhep#1#2#3#4{\href{http://jhep.sissa.it/stdsearch?paper=#2\%28#3\%29#4}{J. High Energy Phys. {\xbold #1#2} ({\xold#3}) {\xold#4}}}
\def\AP#1#2#3{Ann. Phys. {\xbold#1} ({\xold#2}) {\xold#3}}

\def\CMP#1#2#3{Commun. Math. Phys. {\xbold#1} ({\xold#2}) {\xold#3}}
\def\CQG#1#2#3{Class. Quantum Grav. {\xbold#1} ({\xold#2}) {\xold#3}}
\def\IJMPA#1#2#3{Int. J. Mod. Phys. {\xbf A}{\xbold#1} ({\xold#2}) {\xold#3}}

\def\NPB#1#2#3{Nucl. Phys. {\xbf B}{\xbold#1} ({\xold#2}) {\xold#3}}

\def\PLB#1#2#3{Phys. Lett. {\xbf B}{\xbold#1} ({\xold#2}) {\xold#3}}

\newcount\sectioncount
\sectioncount=0
\def\section#1#2{\global\eqcount=0
	\global\subsectioncount=0
        \global\advance\sectioncount by 1
	\ifnum\sectioncount>1
	        \vskip2\baselineskip
	\fi
	\noindent
        \line{\twelvecp\the\sectioncount. #2\hfill}
		\vskip.8\baselineskip\noindent
        \xdef#1{{\old\the\sectioncount}}}
\newcount\subsectioncount
\def\subsection#1#2{\global\advance\subsectioncount by 1
	\vskip.8\baselineskip\noindent
	\line{\tencp\the\sectioncount.\the\subsectioncount. #2\hfill}
	\vskip.5\baselineskip\noindent
	\xdef#1{{\old\the\sectioncount}.{\old\the\subsectioncount}}}
\newcount\appendixcount
\appendixcount=0
\def\appendix#1{\global\eqcount=0
        \global\advance\appendixcount by 1
        \vskip2\baselineskip\noindent
        \ifnum\the\appendixcount=1
        \hbox{\twelvecp Appendix A: #1\hfill}\vskip\baselineskip\noindent\fi
    \ifnum\the\appendixcount=2
        \hbox{\twelvecp Appendix B: #1\hfill}\vskip\baselineskip\noindent\fi
    \ifnum\the\appendixcount=3
        \hbox{\twelvecp Appendix C: #1\hfill}\vskip\baselineskip\noindent\fi}
\def\acknowledgements{\vskip2\baselineskip\noindent
        \underbar{\it Acknowledgements:}\ }
\newcount\eqcount
\eqcount=0
\def\Eqn#1{\global\advance\eqcount by 1
\ifnum\the\sectioncount=0
	\xdef#1{{\old\the\eqcount}}
	\eqno({\oldstyle\the\eqcount})
\else
        \ifnum\the\appendixcount=0
	        \xdef#1{{\old\the\sectioncount}.{\old\the\eqcount}}
                \eqno({\oldstyle\the\sectioncount}.{\oldstyle\the\eqcount})\fi
        \ifnum\the\appendixcount=1
	        \xdef#1{{\oldstyle A}.{\old\the\eqcount}}
                \eqno({\oldstyle A}.{\oldstyle\the\eqcount})\fi
        \ifnum\the\appendixcount=2
	        \xdef#1{{\oldstyle B}.{\old\the\eqcount}}
                \eqno({\oldstyle B}.{\oldstyle\the\eqcount})\fi
        \ifnum\the\appendixcount=3
	        \xdef#1{{\oldstyle C}.{\old\the\eqcount}}
                \eqno({\oldstyle C}.{\oldstyle\the\eqcount})\fi
\fi}
\def\eqn{\global\advance\eqcount by 1
\ifnum\the\sectioncount=0
	\eqno({\oldstyle\the\eqcount})
\else
        \ifnum\the\appendixcount=0
                \eqno({\oldstyle\the\sectioncount}.{\oldstyle\the\eqcount})\fi
        \ifnum\the\appendixcount=1
                \eqno({\oldstyle A}.{\oldstyle\the\eqcount})\fi
        \ifnum\the\appendixcount=2
                \eqno({\oldstyle B}.{\oldstyle\the\eqcount})\fi
        \ifnum\the\appendixcount=3
                \eqno({\oldstyle C}.{\oldstyle\the\eqcount})\fi
\fi}
\def\multi{\global\advance\eqcount by 1}
\def\multieq#1#2{\xdef#1{{\old\the\eqcount#2}}
        \eqno{({\oldstyle\the\eqcount#2})}}
\newtoks\url
\def\Href#1#2{\catcode`\#=12\url={#1}\catcode`\#=\active#2}
\def\href#1#2{{#2}}

\parskip=3.5pt plus .3pt minus .3pt
\baselineskip=14pt plus .1pt minus .05pt
\lineskip=.5pt plus .05pt minus .05pt
\lineskiplimit=.5pt
\abovedisplayskip=18pt plus 4pt minus 2pt
\belowdisplayskip=\abovedisplayskip
\hsize=14cm
\vsize=19cm
\hoffset=1.5cm
\voffset=1.8cm
\frenchspacing
\footline={}
\raggedbottom

\def\sss{\scriptscriptstyle}
\def\*{\partial}
\def\punkt{\,\,.}
\def\komma{\,\,,}

\def\={\!=\!}
\def\small#1{{\hbox{$#1$}}}
\def\half{\small{1\over2}}
\def\fraction#1{\small{1\over#1}}
\def\fr{\fraction}
\def\Fraction#1#2{\small{#1\over#2}}
\def\Fr{\Fraction}

\def\eg{{\tenit e.g.}}

\def\ie{{\tenit i.e.}}
\def\etal{{\tenit et al.}}

\def\a{\alpha}
\def\b{\beta}

\def\d{\delta}
\def\e{\varepsilon}
\def\g{\gamma}

\def\G{\Gamma}

\def\O{\Omega}

\def\w{\!\wedge\!}

\def\Tr{\hbox{Tr}\,}

\def\D{{\cal D}}
\def\EE{{\cal E}}

\def\AA{{\cal A}}

\def\dd#1{{\partial\over\partial#1}}


\headtext={Bao, Bengtsson, Cederwall, Nilsson: ``Membranes for
Topological M-Theory''}

\line{
\epsfysize=20mm
\epsffile{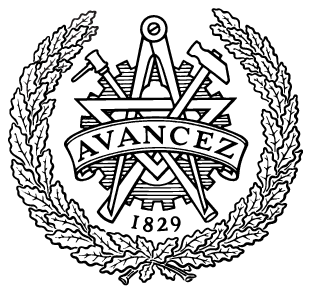}
\hfill}
\vskip-17mm
\line{\hfill G\"oteborg preprint}
\line{\hfill hep-th/0507077}
\line{\hfill July, {\old2005}}
\line{\hrulefill}

\vfill

\centerline{\sixteenhelvbold Membranes for Topological M-Theory}

\vfill

\centerline{\twelvehelvbold Ling Bao, Viktor Bengtsson, Martin
  Cederwall and Bengt E.W. Nilsson}

\vfill

\centerline{\it Fundamental Physics}
\centerline{\it Chalmers University of Technology }
\centerline{\it S-412 96 G\"oteborg, Sweden}

\vfill

{\narrower\noindent 
\underbar{Abstract:} 
We formulate a theory of topological membranes on manifolds with
$G_2$ holonomy. The BRST charges of the theories are 
the superspace Killing vectors (the generators of global supersymmetry)
on the background with reduced holonomy $G_2\subset Spin(7)$. 
In the absence of spinning formulations
of supermembranes, the starting point is an $N=2$ target space supersymmetric
membrane in seven euclidean dimensions. The reduction of the
holonomy group implies a twisting of the rotations in the tangent
bundle of the branes with ``R-symmetry'' rotations in the normal
bundle, in contrast to the ordinary spinning formulation of
topological strings, where twisting is performed with internal
$U(1)$ currents of the $N=(2,2)$ superconformal algebra.
The double dimensional reduction on a circle of the topological
membrane gives the strings of the topological A-model (a by-product of
this reduction is a Green--Schwarz formulation of topological strings). 
We conclude that the action is BRST-exact modulo topological terms and
fermionic equations of motion.
We discuss the r\^ole of topological membranes in topological M-theory and
the relation of our work to recent work by Hitchin and by Dijkgraaf
{\it et al}. 
\smallskip}
\vfill

\font\xxtt=cmtt6

\vtop{\baselineskip=.6\baselineskip\xxtt
\line{\hrulefill}
\catcode`\@=11
\line{email: ling.bao@fy.chalmers.se, viktor.bengtsson@chalmers.se,
martin.cederwall@chalmers.se, tfebn@fy.chalmers.se\hfill}
\catcode`\@=\active
}

\eject

\ref\Dijkgraafetal{R. Dijkgraaf, S. Gukov, A. Neitzke and C. Vafa,
{\xit ``Topological M-theory as unification of form theories of
gravity''}, \hepth{0411073}.}

\ref\GerasShatashvili{A.A. Gerasimov and S.L. Shatashvili,
{\xit ``Towards integrability of topological strings. I: Three-
                  forms on Calabi-Yau manifolds''},
 \jhep{04}{11}{2004}{074} 
[\hepth{0409238}].}

\ref\deBoerone{J. de Boer, A. Naqvi, and A. Shomer, 
{\xit ``Towards a topological G2 string''},
\hepth{0502140}; {\xit ``The topological G2 string''},
\hepth{0506211}.}


\ref\HitchinGenCY
{N. Hitchin,
{\xit ``Generalized Calabi--Yau manifolds''},
Quart.\ J.\ Math.\ Oxford Ser.\  {\xbf 54} (2003) 281\hfill\break
[\mathdg{0209099}].}

\ref\HitchinThreeformsinsix
{N. Hitchin,
{\xit ``The geometry of three-forms in six and seven dimensions''},
J.\ Diff.\ Geom.\  {\xbf 55} (2000) 547 [\mathdg{0010054}].}

\ref\HitchinStableforms
{N. Hitchin,
{\xit ``Stable forms and special metrics''},
\mathdg{0107101}.}

\ref\HarveyMoore{J. Harvey and G. Moore,
{\xit ``Superpotentials and membrane instantons''},
\hepth{9907026}.}

\ref\WittenTopsigma{E. Witten, {\xit ``Topological sigma models''},
\CMP{118}{1988}{411}.}

\ref\WittenMirror{E. Witten, 
{\xit ``Mirror manifolds and topological field theory''},
\hepth{9112056}.}

\ref\NeitzkeVafa{A. Neitzke and C. Vafa, 
{\xit ``Topological strings and their physical applications''},
\hepth{0410178}.}

\ref\KodairaSpencergravity{M. Bershadsky, S. Cecotti, H. Ooguri, and C. Vafa,
{\xit ``Kodaira--Spencer theory of gravity and exact results for
quantum string amplitudes''},  
\CMP{165}{1994}{311}, [\hepth{9309140}].}

\ref\Kahlergravity{M. Bershadsky and V. Sadov,
{\xit ``Theory of K\"ahler gravity''}, \IJMPA{11}{1996}{4689} 
\hfill\break[\hepth{9410011}].}

\ref\Dirichletthreebrane{M. Cederwall, A. von Gussich, B.E.W. Nilsson
and A. Westerberg, 
{\xit ``The Dirichlet super-three-brane in ten-dimensional type IIB
supergravity''}, 
\NPB{490}{1997}{163} [\hepth{9610148}].}

\ref\Dirichletpbrane{M. Cederwall, A. von Gussich, B.E.W. Nilsson, P. Sundell
 and A. Westerberg,
{\xit ``The Dirichlet super-p-brane in ten-dimensional type IIA and
IIB supergravity''}, 
\NPB{490}{1997}{179} [\hepth{9611159}].}

\ref\PestunWitten{V. Pestun and E. Witten, 
{\xit ``The Hitchin functionals and the topological B-model at one loop''},
\hfill\break\hepth{0503083}.}

\ref\BeckerBeckerStrominger{K. Becker, M. Becker and A. Strominger,
{\xit ``Five-branes, membranes and non-perturbative
 string theory''}, \NPB{456}{1995}{130} [\hepth{9507158}].}

\ref\Joyce{D. Joyce, {\xit ``Compact manifolds with special
 holonomy''}, Oxford 2000.} 

\ref\TownsendCalM{P.K. Townsend,
{\xit ``PhreMology: calibrated M-branes''},
\CQG{17}{2000}{1267}
\hfill\break[\hepth{9911154}].}

\ref\GutowskiPapadopAdSCal{J. Gutowski and G. Papadopoulos,
{\xit ``AdS calibrations''},
\PLB{462}{1999}{81}
[\hepth{9902034}].}

\ref\Granaetal{M. Gra\~na, R. Minasian, M. Petrini, and A. Tomasiello,
{\xit ``Supersymmetric backgrounds from generalized Calabi--Yau manifolds''},
\jhep{08}{}{2004}{046} [\hepth{0406137}].}

\ref\BergshoeffetalMembrane{E. Bergshoeff, E. Sezgin and P.K. Townsend,
{\xit ``Properties of the eleven-dimensional supermembrane theory''},
\AP{185}{1988}{330}.}

\ref\BerkovitsMembrane{N. Berkovits, {\xit ``Covariant quantization of the supermembrane''},
 \jhep{02}{09}{2002}{051} 
[\hepth{02011151}].}

\ref\GrassiVanhove{P.A. Grassi and P. Vanhove,
{\xit ``Topological M theory from pure spinor formalism''},
\hfill\break\hepth{0411167}.}

\ref\PiolineNicolaiPlefkaWaldron{B. Pioline, H. Nicolai, J. Plefka and
 A. Waldron, {\xit ``R${}^{\sss4}$ couplings, the fundamental membrane and
 exceptional theta correspondences''}, \jhep{01}{03}{2001}{036} 
[\hepth{0102123}].}

\ref\SuginoVanhove{F. Sugino and P. Vanhove, {\xit ``U-duality from matrix
 membrane partition function''}, \PLB{522}{2001}{145}
 [\hepth{0107145}].}

\ref\Nekrasov{N. Nekrasov, {\xit ``A la recherche de la m-th\'eorie
 perdue --- Z-theory: chasing m/f theory''}, 
 \hepth{0412021}.}

\def\O{\Omega}
\def\Od{{\star}\Omega}
\def\s{\sigma}
\def\sd{{\star}\sigma}

\def\k{\kappa}
\def\m{\mu}

\def\OO{{\Bbb O}}
\def\HH{{\Bbb H}}
\def\CC{{\Bbb C}}
\def\RR{{\Bbb R}}

\def\E{{\scr E}}

\def\dd#1{{\partial\over\partial#1}}

\section\Introduction{Introduction}The notion of topological M-theory
was introduced recently by  
Dijkgraaf \etal\ [\Dijkgraafetal] (see also [\GerasShatashvili]). 
In analogy with topological string theory [\WittenTopsigma,\WittenMirror] 
(for a recent review, see ref. [\NeitzkeVafa]), 
one expects here a 
topological membrane world-volume theory to give rise to a field theory 
in a seven-dimensional target space. In the string case both the world-sheet 
and the six-dimensional target space theories are fairly well understood, 
the latter being in fact string field theories constructed from the
world-sheet BRST charge. 
Although Calabi--Yau three-folds have special properties in this
context [\NeitzkeVafa],  
topological strings exist also on special holonomy manifolds of other
dimensionalities, see 
\eg\ ref. [\deBoerone].
The features found in the topological string case would for many reasons 
be very valuable to understand 
also in the membrane/M-theory case. One important reason is connected
to the r\^ole  
topological string amplitudes play in 
compactification of physical string theories. One may also wonder if a
better understanding 
of topological M-theory may indicate how to approach the problem of finding a 
microscopic formulation of M-theory, possibly including a
 quantisation of the membrane. 

In ref. [\Dijkgraafetal], the authors took a first step towards this
goal by suggesting 
the form of the effective target space field theory of topological M-theory. 
Such an effective theory may be obtained by arguing that the theory
and its topological  
properties should be connected to those of the A and/or B 
topological string models by dimensional reduction in much the same way as the
physical field theories in ten and eleven dimensions are
related. Similarly, one should be  
able to connect the topological string world-sheet theories to the topological 
membrane one by subjecting the latter to a double dimensional reduction.

The target space aspects were discussed in some detail in
ref. [\Dijkgraafetal],  
where the crucial r\^ole of Hitchin functionals
[\HitchinThreeformsinsix,\HitchinStableforms] was  
elaborated upon. These are special functionals of $p$-forms which can
be connected to  
metric fields by some rather complicated non-linear relations. The resulting 
theory was given the 
appropriate name {\it form-gravity} in ref. [\Dijkgraafetal]. 
By starting from a Hitchin 3-form 
on a seven-dimensional $G_2$ 
holonomy manifold, the authors of ref. [\Dijkgraafetal] show that by 
dimensional reduction various well-known topological form-gravity theories in 
lower dimensions are obtained. In particular, one finds 
the Kodaira--Spencer theory [\KodairaSpencergravity] for the complex structure 
deformations in the 
B-model and the K\"ahler gravity theory [\Kahlergravity] of the
K\"ahler deformations in the  
A-model, albeit produced in a particular interacting form.

 At the classical level the connection between form-gravity based 
on a six-dimensional Hitchin functional and the topological B-model
was made explicit by  
relating the corresponding tree-level partition functions to each
other. However at one 
loop level, where the B-model is known to compute a special combination of 
Ray--Singer torsion invariants [\KodairaSpencergravity], 
it was recently demonstrated by Pestun and Witten 
[\PestunWitten] that one needs to use the extended Hitchin functional
introduced in 
ref. [\HitchinGenCY] to obtain the same one-loop partition
function. This connection to the  
extended Hitchin functionals is intriguing since they play a r\^ole also
in flux compactifications [\Granaetal] 
on the generalised Calabi--Yau manifolds 
discussed by Hitchin in his paper. 

The natural next step seems to be to construct a topological 
membrane theory that may be related to the topological M-theory
mentioned above.
That is,  we want to construct a membrane embedded in a
seven-dimensional space with 
$G_2$ holonomy whose  
effective action is the Hitchin functional 
3-form gravity theory discussed in ref. [\Dijkgraafetal]. The usual
approach to derive 
topological strings by means of twisting does not seem to work here since
it is based on the spinning string, or NSR, formulation which is
lacking in the membrane case. 
Here we will instead approach this problem by starting from the
Green--Schwarz (GS)  
formulation of the membrane [\BergshoeffetalMembrane]. Of course, since 
the superstring, and perhaps also the supermembrane, are quantized most easily  
using Berkovits'
pure spinor formulation [\BerkovitsMembrane], this is probably an even more 
suitable starting point. This point was discussed recently also in [\GrassiVanhove].  
We note here
that although the GS formulation of string theory is as standard as the NSR
one,  it does  not seem to have been used yet in the construction of 
topological strings. As will be clear below such a GS formulation will
come out  of the results presented here for the topological membrane. 

One important aspect of twisting in the construction of the topological 
string from a two-dimensional supersymmetric sigma model is that it turns
a spin-$3\over2$ supersymmetry current into a spin-1 object that can be 
interpreted as a BRST current. This kind of twisting is accomplished 
by enforcing the identification of the world-sheet Lorentz symmetry with 
an $so(2)$ R-symmetry giving fermionic quantities unphysical integer 
spin values. In the GS formulation of the membrane, which is the starting 
point in our approach to the topological membrane, such an unphysical 
spin--statistics relation on the world-volume is already in effect since the 
supercoordinates in the target space $(x^{m},\psi^{\hat{\m}I})$ contain the 
anticommuting world-volume scalars $\psi^{\hat{\mu}I}$
(the ranges of the various 
indices will be specified later). For trivial target spaces like 
eleven-dimensional flat space, the gauge-fixing of the $\k$-symmetry
generates an ordinary 
supermultiplet in three dimensions with physical spin fermions
[\BergshoeffetalMembrane].  
However, in the context discussed here no twisting needs to be 
done by hand, a fact that has been 
noticed before in ref. [\HarveyMoore]. As discussed in detail
in section 3, a similar  
phenomenon to twisting does occur but now as an automatic consequence 
of combining $G_2$ holonomy and the 
tangent space symmetry remaining
after the introduction of the membrane into target space. 
This twisting leaves the bosonic and 
fermionic fields in the same representation of the surviving symmetry. 
We will however not fix the gauge, and 
for the most part work with a fully 
 $\k$-symmetric theory with a (1+7)-dimensional parameter.

This paper is organised as follows. In section 2 we start by discussing
the $G_2$ 3-form gravity theory that the topological 
membrane is 
supposed to generate in the seven-dimensional target space. 
Different action functionals 
are  presented for this theory, one of which we believe is new. This section
also describes the supergeometry into which the bosonic seven-dimensional $G_2$
holonomy manifolds can be embedded. The supercoordinates are 
$Z^M=(x^m,\theta^{\hat{\mu}I})$
where $m$ runs over seven values and $\hat{\mu}I$ enumerates two $(I=1,2)$ 
eight-dimensional
spinors $(\hat{\mu}=1,..,8)$. The supergeometry in encoded by a standard 
vielbein (supersiebenbein)
and a superspace 3-form $C_{MNP}(Z)$. The Bianchi identities are discussed
and an explicit 3-form superfield is derived, but only in the flat space limit.
As also explained, the full expansions in fermionic coordinates of the 
curvature dependent 3-form and vielbein 
superfields can be obtained by
a lengthy iterative procedure which we hope to come back to in a
future publication 
(for a similar discussion, see refs. [\Dirichletthreebrane,\Dirichletpbrane]).

In section 3 we discuss the $\k$-symmetric membrane theory that we propose
as the starting point for deriving a topological membrane. The r\^ole
of $G_2$ in obtaining the BRST charge from a partially gauge fixed 
world-volume action is explained and arguments indicating the 
topological nature of the action,
namely the fact that it is BRST exact, are presented. This discussion
is carried out in the full theory but the calculation of the action is
performed 
only to lowest order in the curvature and a full proof will require 
more work.

In the concluding section 4 we make a few additional remarks and comments.
Properties of the octonions are used heavily in this paper and some aspects 
can be found in the appendices. In appendix 
A we discuss $G_2$ tensors, projection operators and the relation to
quaternions, while  
in appendix B we give the explicit form of the flat superspace
3-form based on the octonionic structure constants.

\section\GTwoHolonomy{$G_2$ Holonomy}Seven-dimensional manifolds with
$G_2$ holonomy have special properties, among which are Ricci-flatness
and a single covariantly constant (Killing) spinor.  

When
holonomy is restricted to lie in a $G_2$ subgroup, a (partial) gauge
choice can be made for the spin connection to make it lie entirely in
the Lie algebra $G_2$. Then, $G_2$ singlets can be defined as constant
over the manifold, and this thus applies to special elements of 
any $Spin(7)$ representation containing a $G_2$
singlet. So, there is a constant spinor, since $8\rightarrow1\oplus7$,
and a constant 3-form $\Omega$, since $35\rightarrow1\oplus7\oplus27$. In a
flat frame the 3-form may be chosen as $\Omega_{abc}=\s_{abc}$, the
octonionic structure constants, invariant under the action of $G_2$, 
the automorphism group of $\OO$ (see Appendix A for details). 

\subsection\GTwoAndThreeForm{The 3-form}Hitchin
[\HitchinThreeformsinsix,\HitchinStableforms] has 
constructed a model containig a 3-form fields, whose solutions are
$G_2$ manifolds. This is certainly a part of topological M-theory. 
The metric is constructed from the 3-form as
$$
\sqrt gg_{mn}
=-\fr{144}\e^{m_1\ldots m_7}\O_{mm_1m_2}\O_{nm_3m_4}\O_{m_5m_6m_7}
\komma\Eqn\GORelation
$$
Hitchin gives the action $S=\int d^7x K^{1/9}$.
A Polyakov type action, due to Nekrasov [\Nekrasov], giving both the
relation (\GORelation) and the
covariant constancy of the 3-form, is 
$$
S'=\Fr29\int d^7x\left(\sqrt g
-\fr{288}g^{mn}\e^{m_1\ldots
  m_7}\O_{mm_1m_2}\O_{nm_3m_4}\O_{m_5m_6m_7}
\right)\komma\Eqn\AuxActionOne
$$
The metric is auxiliary and determined
by its equation of motion. The constant in front is chosen so that
the action is normalised to the volume. In a frame where (locally)
$\Omega_{abc}=\s_{abc}$, one thus has $g_{ab}=\d_{ab}$, which is checked by
$\s_{acd}\s_{bef}\sd_{cdef}=-24\d_{ab}$ (see Appendix A).

Varying the action (\AuxActionOne) w.r.t. $\O$ gives 
$$
\eqalign{
-\fr{9\times144}\int d^7x\bigl(
&2g^{rn}\e^{stm_3\ldots m_7}\O_{nm_3m_4}\O_{m_5m_6m_7}\cr
&+g^{mn}\e^{m_1\ldots
  m_4rst}\O_{mm_1m_2}\O_{nm_3m_4}\bigr)\d\O_{rst}\punkt\cr
}
\Eqn\OmegaVariation
$$
Using the relations of Appendix A to calculate the two terms, one finds using
the expression for $g_{mn}$
that they both are proportional
to the same expression, and that the variation (\OmegaVariation)
becomes
$$
\fr3\int\Od\wedge\d\O=\fr{18}\int d^7x\sqrt
g\,\O^{mnp}\d\O_{mnp}\punkt
\Eqn\LinearVariation
$$

The relation (\GORelation) for the metric may equivalently be written in the
implicit form
$$
g_{mn}=\fr6g^{p_1q_1}g^{p_2q_2}\O_{mp_1p_2}\O_{nq_1q_2}\komma\Eqn\GOImplicit
$$
which is used by
Hitchin in expressing the variation of his action in the ``linear
form'' (\LinearVariation). This latter relation could as well be
obtained from an action, which now takes a much more conventional
form:
$$
S''=-\fr6\int d^7x\sqrt g\,\bigl
(1-\fr6g^{m_1n_1}g^{m_2n_2}g^{m_3n_3}\O_{m_1m_2m_3}\O_{n_1n_2n_3}\bigr)
\Eqn\NormalAction
$$
(what varying this action w.r.t. $g^{mn}$ really gives is 
$\O_{mpq}\O_n{}^{pq}=-g_{mn}(1-\fr6\O_{pqr}\O^{pqr})$, which after
contracting the free indices with $g^{mn}$ gives
$\O_{pqr}\O^{pqr}=42$, and thus $g_{mn}=\fr6\O_{mpq}\O_n{}^{pq}$).
Variation of the action (\NormalAction) w.r.t. $\O$ gives an
expression proportional to (\LinearVariation) directly, without any
use of the algebraic identities of Appendix A.

The 3-form is part of the geometric background for propagation of
membranes.
The expression ``$\Omega=\s$'' is purely bosonic. In a superspace,
$\Omega$ will contain more components when expressed in flat basis, 
due to torsion (see appendix B). 


\subsection\SuperSpace{Superspace and Supersymmetry}The superspace 
we want to consider has bosonic coordinates which are
the coordinates of a euclidean manifold with $G_2$ holonomy. In
addition there will be fermionic coordinates. These are {\it a priori}
a set of real spinors in the 8-dimensional representation of
$Spin(7)$, but when $Spin(7)\rightarrow G_2$ each spinor decomposes
as $8\rightarrow1\oplus7$. The $\gamma$-matrices of $Spin(7)$ are real
and antisymmetric, so it is clear that an even number of spinors are
needed, together with an internal $Sp(2n)$ in order to have a
non-vanishing torsion. 
We will 
choose the simplest possibility, $n=1$, giving a doublet of spinors,
for reasons that become obvious in the following subsection.
This superspace is obtained from $D=11$ superspace, with twice as many
fermionic coordinates, as a truncation of the $Spin(7)\times
SL(2,\CC)$ subgroup of $Spin(1,10)$ to $Spin(7)\times SL(2,\RR)$,
where the spinors in the representation $32\rightarrow(8,2_\CC)$ are
demanded to be real. 

A convenient realisation is to consider a vector as an imaginary
octonion, $v\in\OO'$, and a
spinor as an arbitrary octonion, $s\in\OO$. Letting the orthonormal basis of
$\OO'$ be $\{e_a\}_{a=1}^7$, multiplication by $\gamma^a$  is
identified with left multiplication of a spinor with $e_a$, \ie, $vs$
is again a spinor. 
The octonionic multiplication table,
$e_ae_b=-\d_{ab}+\s_{abc}e_c$, 
tells us that the real $\gamma$-matrices square to $-1$ (a property 
which will be crucial for supermembranes). 

Before moving on let us fix some notation. Superspace coordinates are written
$$
Z^{M} = (x^{m}, \psi^{\hat{\mu}I})\komma\quad m = 1,\ldots 7\komma 
\quad \hat{\mu} =  0, \ldots, 7\komma\quad I = 1,2\punkt\eqn
$$
Flat indices are written $(a,\hat\a I)$.
The spinor index will often be divided into $(0,\a=1,\ldots,7)$, 
reflecting the decomposition
$8\rightarrow1\oplus7$. This division applies also to curved indices,
as long as one only considers super-diffeomorphisms that leave the
singlet inert, and we use the notation
$$ \psi^{\hat{\mu}I} = (\theta^{I}, \psi^{\mu I})\punkt\eqn$$
Bosonic and fermionic vielbeins are written,
$$ E^{a}\komma\quad \EE^{\hat{\a}I} = (\EE^{I}, \EE^{\a I})\komma\eqn$$
and the purely bosonic vielbein, $e_{m}{}^{a}$. 

The $\gamma$ matrices encoded in the left
multiplication of a spinor $\lambda=\lambda^{\hat\a}e_{\hat\a}$ by an
imaginary unit $e_a$ are
$$
\eqalign{
(\g^a)_{\a\b}&=\s^a{}_{\a\b}\komma\cr
(\g^a)_{0\a}&=\d^a_\a\punkt\cr
}\eqn
$$
They satisfy 
$
\{ \gamma^a,\gamma^b \} = -2\delta^{ab}\komma
$
where the minus sign is necessary for real $\g$-matrices.

The Clifford algebra 
is spanned by the $so(7)$-invariant tensors
$\delta^{\hat{\alpha}}_{\phantom{\hat{\alpha}}\hat{\beta}}$,
$(\gamma^a)^{\hat{\alpha}}_{\phantom{\hat{\alpha}}\hat{\beta}}$,
$(\gamma^{ab})^{\hat{\alpha}}_{\phantom{\hat{\alpha}}\hat{\beta}}$ and
$(\gamma^{abc})^{\hat{\alpha}}_{\phantom{\hat{\alpha}}\hat{\beta}}$,
of which the first and last are symmetric and the second and third
antisymmetric matrices.
The decomposition in terms of $G_2$-invariant tensors is
$$
\eqalign{
\delta^{\hat{\alpha}}_{\phantom{\hat{\alpha}}\hat{\beta}} & = \left[
\matrix{ 1 & 0 \cr
0 & \delta^{\alpha}_{\phantom{\alpha}\beta} \cr
} \right] \cr
(\gamma^a)^{\hat{\alpha}}_{\phantom{\hat{\alpha}}\hat{\beta}} & = \left[
\matrix{ 0 & \delta^{a}_{\phantom{a}\beta} \cr
-\delta^{a\alpha} & \sigma^{a\alpha}_{\phantom{a\alpha}\beta} \cr
} \right] \cr
(\gamma^{ab})^{\hat{\alpha}}_{\phantom{\hat{\alpha}}\hat{\beta}} & = \left[
\matrix{ 0 & -\sigma^{ab}_{\phantom{ab}\beta} \cr
\sigma^{ab\alpha} & -\star\sigma^{ab\alpha}_{\phantom{ab\alpha}\beta} 
- 2\d^{ab}_{\alpha\beta} \cr
} \right] \cr
(\gamma^{abc})^{\hat{\alpha}}_{\phantom{\hat{\alpha}}\hat{\beta}} & = \left[
\matrix{ \sigma^{abc} & -\star\sigma^{abc}_{\phantom{abc}\beta} \cr 
-\star\sigma^{abc\alpha} & 6
\delta_{(\alpha}^{[a}\sigma_{\beta)}^{\phantom{\beta)}bc]} -
\delta^{\alpha}_{\phantom{\alpha}\beta}\sigma^{abc} \cr } 
\right]\punkt
}\eqn
$$

Solving the dimension-0 part of the Bianchi identities reveals a
possible solution in terms of $SO(7)$ $\g$-matrices (a wider class of
solutions in terms of $G_{2}$-invariants exists). A possibility which
becomes a requirement when treating $\k$-symmetry for the membrane. 
We choose $T^a_{\a I,\b J}=2\e_{IJ}(\g^a)_{\a\b}$,
implying 
$$
\eqalign{
T_{\a I,\b J}{}^a&=2\e_{IJ}\s^a{}_{\a\b}\komma\cr
T_{0I,\a J}{}^a&=2\e_{IJ}\d_\a^a\komma\cr
T_{0I,0J}{}^a&=0\punkt\cr
}\eqn
$$
The background will contain a 3-form potential $C$ (descending from
the one in $D=11$) with 4-form field strength, $G$, whose dimension-0 part
is taken to be  
$G_{ab,\hat\a I,\hat\b J}=-2\e_{IJ}(\g_{ab})_{\hat\a\hat\b}$:
$$
\eqalign{
G_{ab,\a I,\b J}&=2\e_{IJ}(2\d_{\a\b}^{ab}+\sd_{ab\a\b})\komma\cr
G_{ab,0I,\b J}&=2\e_{IJ}\s_{ab\b}\komma\cr
G_{ab,0I,0J}&=0\punkt\cr
}\Eqn\GExpression
$$

The Fierz identity in $D=7$ ensuring the Bianchi identity for $G$ is 
$$
\Yboxdim3pt
(\g^b)_{\hat\a\hat\b}(\g_{ab})_{\hat\g\hat\d}\vert_{\yng(2,2)}=0\komma\eqn
$$ 
where the Young tableau indicates the symmetry structure of the spinor
indices. The expression
$(\g^b)_{\hat\a\hat\b}(\g_{ab})_{\hat\g\hat\d}$ contains only terms
that are antisymmetric in at least three spinor indices, implying that 
$\e_{IJ}\e_{KL}(\g^b)_{\hat\a\hat\b}(\g_{ab})_{\hat\g\hat\d}$
completely symmetrised in the four composite indices $(\hat\a I,\hat\b J,\hat\g
K,\hat\d L)$ vanishes.

The potential $C$, which will be the field that the supermembranes
couples minimally to, is {\it a priori} thought of as a 3-form with
vanishing cohomology class, so that, modulo gauge transformations,
$C_{abc}=0$. Of course, changing $C$ to $C^{(k)}=C+k\Omega$ leaves the
field strength invariant.

The constraints for torsion and field strength used are standard, and
the ones obtained by reduction from $D=11$ and truncation to real
fermions. In
order to use them to extract an explicit form for the dynamics of the
supermembrane introduced in the following section, one would need to
solve these constraints explicitly for the vielbeins and components of
$C$ in terms of the bosonic and fermionic coordinates. This has not
been done, except for in the case of flat manifolds (orbifolds of
tori). In principle, this can be done order by order in the fermions,
and we will indicate how this expansion starts.

The target space coordinates\foot\dagger{The identification of part of the
spinor as vectors involves gauge-fixing all except the bosonic
diffeomorphisms.} are $x^m$, $\psi^{mI}$ and $\theta^I$ . Under (bosonic)
diffeomorphisms, $\delta_\chi x^m=\chi^m$,
$\delta_\chi\psi^{mI}=\psi^{nI}\partial_n\chi^m$, $\delta_\chi\theta^I=0$. 
This means that the derivatives and dual differentials that transform
covariantly are
$$
\matrix{dx^m\hfill&\D_m=\partial_m-\G_{mn}^p\psi^{nI}\dd{\psi^{pI}}\hfill\cr
D\psi^{mI}=d\psi^{mI}+dx^n\psi^{pI}\G^m_{np}\hfill\qquad
            &\dd{\psi^{mI}}\hfill\cr
d\theta^I\hfill&{\partial\over\partial\theta^I}\hfill\cr}\eqn
$$
(if we have differentials that transform covariantly, we can just
contract them with $e_m{}^a$ to get something that is invariant).
In order to reproduce the dimension-0 torsion, the vielbeins are
constructed from the covariant differentials as
$$
\eqalign{
E^a&=(dx^m+\e_{IJ}\Omega^m{}_{np}D\psi^{nI}\psi^{pJ}
+2\e_{IJ}d\theta^I\psi^{mJ})e_m{}^a+\ldots\komma\cr
\EE^{aI}&=D\psi^{mI}e_m{}^a+\ldots\komma\cr
\EE^I&=d\theta^I\punkt\cr
}\Eqn\FormOfVielbeins
$$
We also let $\omega=dx^m\omega_m(x)$ and $D=d+\omega$. 
These terms generate torsion, however, which contains the Riemann
 tensor
 (${\cal T}^{aI}\equiv T^{\a I}\delta_\a^a$):
$$
\eqalign{
T^a&=\e_{IJ}(\Omega^m{}_{np}D\psi^{nI}\wedge D\psi^{pJ}
       +2d\theta^I\wedge D\psi^{mJ})e_m{}^a
      +\Omega^m{}_{np}R_q{}^n\psi^{qI}\psi^{pJ}e_m{}^a\komma\cr
{\cal T}^{aI}&=\psi^{nI}R_n{}^me_m{}^a\komma\cr
}\eqn
$$
where $D\Omega=0$ has been used. The curvature enters with
$R_m{}^n\equiv\half dx^p\w dx^qR_{qp\,m}{}^n$. So, while the correct
torsion terms are generated, the curvature-dependent ones have to be
compensated for by adding terms of higher order in fermions in the
vielbeins. Note, however, that this does not apply to the coefficients
of $d\theta^I$,
which is a $G_2$ singlet, hence not affected by the spin connection,
and furthermore exact.

\subsection\GTwoAndSymmetry{$G_2$ manifolds and supersymmetry}The
existence 
of a constant spinor allows for a Killing spinor, a
fermionic ``isometry'' of the superspace, \ie, a global
supersymmetry. In our superspace with an internal $SL(2)$ index, there
will be a doublet of supersymmetries. We choose a parametrisation
where the superspace Killing vectors, \ie, the supersymmetry
generators are  
$$
Q_I=\dd{\theta^I}\komma\Eqn\QExpression
$$
which obviously fulfill 
$$
\{Q_I,Q_J\}=0\punkt\Eqn\QSquareIsZero
$$
All vielbeins in eq. (\FormOfVielbeins) are invariant under $Q_I$. We
may remark that the simple form (\QExpression) of the supersymmetry
generators depends on the the form of the bosonic vielbeins. If the
fermion bilinears in the bosonic vielbein had been chosen to contain
$\e_{IJ}(d\theta^I\psi^{mJ}-\theta^Id\psi^{mJ})$ instead of 
$2\e_{IJ}d\theta^I\psi^{mJ}$ (which to lowest order corresponds to a
change of bosonic coordinate), one would also have had a term 
$-\e_{IJ}\psi^{mI}{\partial\over\partial x^m}$ in
$Q_I$. Diffeomorphism covariance would then demand that 
${\partial\over\partial x^m}$ is replaced by $\D_m$, so also $\psi$
is transformed. It turns out (by trial and error) that it is
impossible to construct a supersymmetry doublet that starts out this
way and fulfills the nilpotency relation (\QSquareIsZero), due to
curvature terms, so we are left with the choice of eq. (\QExpression).

In the discussion on a topological theory of membranes below, the $Q_I$'s
are the nilpotent operators that will be promoted to BRST operators.

\vfill\eject

\section\TopologicalMembranes{Topological Membranes}In this section,
we will describe in detail how we obtain a topological membrane by
imposing a supersymmetry constraint on supermembranes embedded in a
superspace extending a manifold of $G_2$ holonomy. First we will
introduce 
supermembranes, and investigate the structure of
$\kappa$-symmetry in the background at hand. We proceed to promote the
global supersymmetry generators to BRST operators, thereby turning the
theory into a topological theory. We show that the action, modulo
topological terms and fermionic equations of motion, is not only
BRST-invariant, but also BRST-exact.

\subsection\Supermembranes{Supermembranes on $G_2$ manifolds}A 
supermembrane in seven dimensions should have $N=2$ supersymmetry, \ie,
propagate in a background superspace with two fermionic spinorial
coordinates $\psi^{\hat\mu I}$. Then the four transverse bosons match the
fermions in number, with $\kappa$-symmetry and equations of motion
taken into account. Both bosons and
fermions are {\it a priori} scalars on the world-volume. This can of
course change after some gauge-fixing, \eg\ choosing a static
gauge. The superspace we choose for the propagation of the membrane is
thus taken to be the one described in the previous section. 

When formulating a theory of topological {\it strings} it is
convenient to start from the action of a spinning (world-sheet
supersymmetric) string. For a membrane, no such formulation exists
that is equivalent to the space-time supersymmetric one. Since we
want our membrane to describe part of M-theory, we seem to be forced
to use the ordinary supermembrane action.
The generic action for a supersymmetric membrane is
$$
S=\int d^3\xi\sqrt g+\int C\punkt\Eqn\MembraneAction
$$
where $g$ and $C$ are pullbacks from target superspace to the world-volume.

The 7-dimensional R-symmetry is $SL(2)$. R-symmetry is typically
something one wants to use in a topological twist, but the real forms of
R-symmetry and local world-volume rotations $su(2)$ do not match.
On the other hand, once one decomposes rotations into longitudinal
and transverse, there are lots of $su(2)$'s. When 
$so(7)\rightarrow so(3)\oplus so(4)\approx su(2)\oplus su(2)\oplus
su(2)$,
$7\rightarrow(1,2,2)\oplus(3,1,1)$ and
$8\rightarrow(2,1,2)\oplus(2,2,1)$.
But if we also have the breaking $so(7)\rightarrow G_2$,
$7\rightarrow7$, $8\rightarrow1\oplus7$, we have to consider the
maximal unbroken subalgebra contained in both $G_2$ and $su(2)^3$. In
the case that the embedding of the membrane world-volume is
associative, \ie, if $\sd_{ijka}=0$, or equivalently
$\s_{ijk}=\pm\e_{ijk}$, this is $su(2)\oplus su(2)$, which is a 
maximal subalgebra of $G_2$, and where the second $su(2)$ is the last
of the three in $so(7)\rightarrow su(2)\oplus su(2)\oplus su(2)$ and
the first is the diagonal subalgebra of the first two (this is shown
in detail in appendix A, using the splitting of an octonion into a pair
of quaternions).
For a more general embedding, the same representations are obtained in
a static gauge based on coordinate directions spanning a quaternion.

From a 3-dimensional perspective, we have (before $G_2$ is imposed)
scalars transforming as vectors under R-symmetry $so(4)$, $\phi\in(1,2,2$), and
spinors transforming as either of the chiralities of $so(4)$, 
$\psi\in(2,2,1)$ and/or $\psi'\in(2,1,2)$.
{\it Introduction of $G_2$ implies a twist of one of the spinor
representations}, since it identifies one of the two R-symmetry
$su(2)$'s with the $su(2)$ of space rotations.
This twisting has been observed earlier in ref. [\HarveyMoore].

The lesson from the behaviour of the representations and the effective
twisting is that when one wants to formulate a topological membrane
theory, no twisting ``by hand'' is needed---it is automatically
provided in a space-time supersymmetric formulation.

\subsection\FermionicSymmetries{Fermionic Symmetries}The supermembrane 
action is invariant under global supersymmetry as
well as $\k$-symmetry. Let us discuss these symmetries in some more
detail, beginning with supersymmetry, generated by the vector fields
$Q_\e=\e^I{\partial\over\partial\theta^I}$, with constant parameters $\e^I$. 

All vielbeins, both the bosonic ones $E^a$ and the fermionic ones
$\E^{\hat\a I}=(d\theta^I,\E^{\a I})$, are invariant under
supersymmetry---this is just the statement that supersymmetry is an
isometry of superspace. 
This accounts for the invariance of the
kinetic volume term in the supermembrane action. 

Invariance of the
Wess--Zumino term $\int C$ is guaranteed by the invariance of the
field strength $G$ of eq. (\GExpression). 
The field strength is expressible as constant coefficients times wedge
products of vielbeins, and thus invariant. This implies that the
supersymmetry transformation of $C$ is a total derivative,
$Q_\e C=\e^Id\Lambda_I$. It is indeed possible to choose a
gauge where a stronger statement, namely local invariance, $Q_\e C=0$,
holds. We have constructed $C$ explicitly in such gauges (to lowest
order in curvature), see
appendix B.
The fact that $C$ can be chosen to be completely independent of
$\theta^I$ will later, when $Q_I$ are used as BRST operators, 
be a crucial property.

To begin our expos\'{e} of $\k$-symmetry for the topological membrane
we recount some well known facts concerning the inner workings of said 
symmetry. In order to reduce clutter we drop the $sl(2)$-indices temporarily, 
reinserting them when returning to the topological membrane. We begin by 
introducing the superspace vector field,
$$
\k=\k^{M} \partial_{M} = \k^{\alpha} \E_{\alpha}^{M} \partial_{M}\quad
(\k^{a} = 0)\komma\eqn
$$
the action of which transforms the pullback of a superspace form as,
$$
\delta_{\k}(f^{*} \Omega) = f^{*} { \cal L_{\k} } \Omega  
= f^{*} ( i_{\k}d + d i_{\k}) \Omega\komma\eqn
$$
where $f^{*}$ is a pullback and $ \cal L$ a Lie derivative. 
From here on we will not write out pullbacks explicitly. 
The action of this vector field on the Wess--Zumino term then follows,
$$
\delta_{\k} \int C = \int (i_{\k} d + d i_{\k}) C 
= \int (i_{\k} G + d i_{\k} C) = \int i_{\k} G\komma\eqn
$$ 
and the variation of the vielbein,
$$
\eqalign{
\delta_{\k} E^{A} &= i_{\k} (T^{A} - E^{B} \wedge \omega_{B}{}^{A}) 
+ D i_{\k} E^{A} - (i_{\k} E^{B}) \wedge \omega_{B}{}^{A}\cr 
&= i_{\k} T^{A} - E^{B} \wedge i_{\k} \omega_{B}{}^{A} + D i_{\k}
E^{A}\punkt\cr
}\eqn
$$
By adding a local Lorentz transformation with parameter $i_{\k}
\omega_{B}{}^{A}$, we can reduce the expression to
$
\delta_{\k} E^{A} = i_{\k} T^{A} + D i_{\k} E^{A}
$,
and furthermore, by considering the relevant part of this expression, to
$$
\delta_{\k} E^{a} = i_{\k} T^{a}\punkt\eqn
$$
The variation of the kinetic term then becomes (with pullbacks written out)
$$
\delta_{\k} \sqrt{g} = \Fr12 \sqrt{g} g^{ij} \delta_{\k} g_{ij} 
= \sqrt{g} g^{ij} E_{(i}^{a} E_{j)}^{B} \k^{\a} T_{\a B}{}^{a}\komma\eqn
$$
where we have used
$
\delta_{\k} g_{ij} = \delta_{\k} (E_{i}^{a} E_{j}^{a}) 
= 2 E_{(i}^{a} E_{j)}^{B} \k^{\a} T_{\a B}{}^{a}
$.
At the level of (length-)dimension $0$ this term varies as
$$
\delta_{\k} \sqrt{g} 
= \sqrt{g} g^{ij} E_{i}^{a} E_{j}^{\b} \k^{\a} T_{\a \b}{}^{a} 
= \sqrt{g} E_{j}^{\b} T_{\a \b}{}^{j} \k^{\a}\komma\eqn
$$
whereupon the action consequently transforms as
$$
\delta_{\k} (\int d^{3} \xi \sqrt{g} + \int C) 
= \int d^{3}\xi \sqrt{g} (E_{i}^{\b} T_{\a \b}{}^{i} \k^{\a} 
+ \Fr12 \Fr{\varepsilon^{ijk}}{\sqrt{g}}
E_{k}^{\b} \k^{\a}G_{ij \a \b}).
\punkt\eqn
$$
Turning presently to the case of the $G_{2}$-membrane this
transformation, 
after insertion of the $8$-dim. torsion and field strength, looks like,
$$
\eqalign{
\delta_{\k} S &= \int d^{3}\xi \sqrt{g} (E_{i}^{\hat{\a}I}
T_{\hat{\a}I, \hat{\b}J}{}^{i} \k^{\hat{\b}J}
+ \fr2 \Fr{\varepsilon^{ijk}}{\sqrt{g}} E_{k}^{\hat{\a}I}
G_{ij \hat{\a}I, \hat{\b}J} \k^{\hat{\b}J})\cr 
&=  \int d^{3}\xi \sqrt{g} E_{i}^{\hat{\a}I} (2
(\g^{i})_{\hat{\a} \hat{\b}} - \Fr{\varepsilon^{ijk}}{\sqrt{g}}
(\g_{jk})_{\hat{a} \hat{\b}}) \k^{\hat{\b}J} \varepsilon_{IJ}\komma\cr
}\eqn
$$
which can be rewritten as
$$
\delta_{\k} S = 2 \int d^{3}\xi 
\sqrt{g} E_{i}^{\hat{\a}I} (\g^{i}\Pi_{+})_{\hat{\a} \hat{\b}} 
\k^{\hat{\b}J} \varepsilon_{IJ}\eqn
$$
The $\kappa$-symmetry condition is thus 
$
(\Pi_{+})^{\hat{\alpha}}_{\phantom{\hat{\alpha}}\hat{\beta}}
\kappa^{\hat{\beta}I} = 0,
$
where
$$
(\Pi_{+})^{\hat{\alpha}}_{\phantom{\hat{\alpha}}\hat{\beta}}
\equiv \Fr{1}{2}\left\{
\delta^{\hat{\alpha}}_{\phantom{\hat{\alpha}}\hat{\beta}} +
\Fr{1}{6\sqrt{g}}\varepsilon^{ijk}
(\gamma_{ijk})^{\hat{\alpha}}_{\phantom{\hat{\alpha}}\hat{\beta}}
\right\}\eqn
$$
is the operator which annihilates an infinitesimal
$\kappa$-variation. The fact that
$$
\Gamma^{\hat{\alpha}}_{\phantom{\hat{\alpha}}\hat{\beta}} \equiv
\Fr{1}{6\sqrt{g}}\varepsilon^{ijk}
(\gamma_{ijk})^{\hat{\alpha}}_{\phantom{\hat{\alpha}}\hat{\beta}} =
\Fr{1}{6\sqrt{g}}\varepsilon^{ijk} \left[ 
\matrix{ \sigma_{ijk} & -\star\sigma_{ijk\beta} \cr 
-\star\sigma_{ijk}^{\phantom{ijk}\alpha} & 6\left(
\delta^{(\alpha}_{[i}\sigma^{\beta)}_{\phantom{\beta)}jk]} -
\fr{6}\delta^{\alpha\beta}\sigma_{ijk} \right) \cr } 
\right]\Eqn\GammaExpression
$$
fulfills the conditions $\Tr(\Gamma) = 0$ and $\Gamma^2 = 1$ implies
that $\Pi_{+}$ is a projection operator\foot\dagger{An essential
observation for the working of {\eightmath\char'24}-symmetry 
is that the euclidean
signature of the world-volume is compensated by the fact that the 
gamma matrices square to
minus one. Compared to 11-dimensional Minkowski space there are two
changes of sign. Had only one of
these changes occurred, idempotent projection matrices could not have
been constructed.}. It is then obvious that the
$\kappa$-symmetry condition can be solved by $\kappa =
\Pi_{-}\xi$, where $\Pi_{-}$ is defined as 
$ (\Pi_{-})^{\hat\alpha}{}_{\hat\beta}
\equiv \Fr{1}{2}(
\delta^{\hat\alpha}{}_{\hat\beta} - \Gamma^{\hat\alpha}{}_{\hat\beta} )
$
and $\xi$ is an arbitrary spinor. Since $\Pi_{-}$ projects out half of
the degrees of freedom of $\xi$, $\kappa$ is parametrised by two
scalars and two world-volume vectors
$\{\lambda^{0I},\lambda^{iI}\}$. It can be shown that $\Pi_{\pm}$ are
the only projection operators, which project out precisely half of the
spinors, that can be formed using the $G_2$ invariant tensors only,
and hence we have found the most general $\kappa$-variation. 

A byproduct of the above calculation is that the fermionic equations
of motion are $\Pi_+\g^i\E_i^{\hat\a I}=0$.

A general background will of course contain fermionic excitations,
demanding that $\k$-symmetry is checked also at dimension
$1\over2$. In the present context, however, we are only interested in
superspaces extending any bosonic manifold of $G_2$ holonomy. We do
not consider deformations of the geometry. In topological M-theory,
such deformations should be parametrised by solutions of the Hitchin
model, and purely bosonic.

The algebra of $\kappa$-symmetry is obtained by commuting
$\kappa$-variations of a fermionic variable, 
which after some calculation, mainly involving
transformation of the projection matrix, yields
$$
\eqalign{[\delta_{\tilde{\kappa}},\delta_{\kappa}]\psi^{\hat{\alpha}I} 
= &\e_{LK}(\Pi_{+}\gamma^{i}{\cal E}_i)^{\hat{\alpha}K}
\tilde{\kappa}^{(L}\kappa^{I)} 
- \e_{LK}
(\gamma^{j})^{\hat{\alpha}}_{\phantom{\hat{\alpha}}\hat{\beta}}
(\Pi_{+}\gamma^{i}{\cal E}_i)^{\hat{\beta}K}
\tilde{\kappa}^{[L}\gamma_{j}\kappa^{I]} \cr
& + (\Pi_{-})^{\hat{\alpha}}_{\phantom{\hat{\alpha}}\hat{\beta}} 
\left\{\Fr{1}{2}[ (\delta_{\tilde{\kappa}}\Pi_{-})\xi 
- (\delta_{\kappa}\Pi_{-})\tilde{\xi} ]^{\hat{\beta}I} 
+ ({\cal E}_i)^{\hat{\beta}I}\e_{KL} \tilde{\kappa}^{K}
\gamma^{i}\kappa^{L} \right\} \cr& - ({\cal
E}_i)^{\hat{\alpha}I} \e_{KL}\tilde{\kappa}^{K}\gamma^{i}\kappa^{L}\punkt
}\Eqn\KappaAlgebra
$$
It is straight-forward to see that the three rows represent
fermionic equations of motion, $\kappa$-transformations and world-volume
diffeomorphisms, respectively. This is the point where it becomes
clear that the formulation, due to the mismatch between fermions and
bosons off-shell, is an on-shell formulation---part of the gauge
symmetry only works modulo equations of motion.

Although we will not develop on this in the present paper, it is worth
mentioning that $\k$-symmetry can in fact be treated in a completely
covariant manner on a $G_2$ manifold. The projection $\k=\Pi_+\k$ may
be solved by parametrising $\k$ in terms of a scalar and a
world-volume vector as
$$
\eqalign{
\k^0&=(1-y)\xi\komma\cr
\k^\a&=z^\a\xi+(E_i^\a-\fr{2\sqrt g}\e_i{}^{jk}\s^\a{}_{jk})\zeta^i\komma\cr
}\Eqn\CovariantKappa
$$
where $y={1\over6\sqrt g}\e^{ijk}\s_{ijk}$, 
$z_\a={1\over6\sqrt g}\e^{ijk}\sd_{ijk\a}$.
In a situation where the scalar part has been fixed, the remaining
gauge symmetry (closing on-shell) will be a super-diffeomorphism
algebra with an $SL(2)$ doublet of world-volume vectors as fermionic
generators.

There is an interplay between the global supersymmetry and the local
$\k$-symmetry, in the sense that both transform the singlet fermions
$\theta^I$. Even if the supersymmetry generators obey eq. (\QSquareIsZero)
exactly and without reference to the embedding of the membrane
world-volume, this ceased to be true once $\k$-symmetry is gauge
fixed. When some gauge is chosen that involves $\theta^I$ (which any
gauge has to), compensating gauge transformations have to be
introduced in order that the redefined supersymmetry generator
transforms within the constraint surface defined by the gauge
choice. Then, due to the commutation relation (\KappaAlgebra), the
nilpotency relation (\QSquareIsZero) only holds on-shell, \ie, modulo
fermionic equations of motion.

\subsection\TopologicalMembranes{Topological Membranes}In order 
to restrict the supermembrane theory to a topological theory,
we want to promote the two supercharges $Q_I$ to BRST operators, and
let the theory be defined by cohomology of these. Unlike the theory of
topological strings, where one in a conformal gauge has a split in
left- and right-movers (or holomorphic and anti-holomorphic dependence
of the world-sheet coordinate), there is no such natural split, and
one has to treat the two supersymmetry generators simultaneously.

We have already shown how the invariance of the supermembrane action
works. If the theory is to become a cohomological field theory, it is
important that the action not only is invariant, but also trivial in
cohomology, \ie, BRST-exact. This means that there should exist a
functional $\Sigma^I$ with 
$$
Q_I\Sigma^J=\d^J_IS\punkt\Eqn\ExactAction
$$
The simple form of the supersymmetry generators assures that this is
achieved by $\Sigma^I=\int{\cal L}\theta^I$, if $Q_I{\cal L}=0$
locally on the world-volume, and the action is then invariant without
resort to partial integration. We have demonstrated earlier that
this is actually the case, due to the fact that a gauge can be chosen
where the 3-form $C$ is independent of $\theta$. The proof of this
statement involved the explicit construction of the superspace 3-form,
which used flat space expressions, but should be possible to
generalise.

The ``pre-action'' $\Sigma^I$ is defined modulo $Q$-exact terms,
encoded in $Q_I\Xi^{JK}=\d_I^K\Delta\Sigma_{\mathstrut}^J$, 
which can be seen as ``gauge transformations'' in the complex.
It is important that other gauge symmetries in the model are
consistent with this one, in the sense that $\Sigma^I$ must be
invariant modulo terms of this trivial type. This applies especially
to $\k$-symmetry, which is not manifest. Indeed, the fact that the
$\k$-variation of ${\cal L}$  is a total derivative ensures that, with
the above form of $\Sigma^I$, $\d_\k\Sigma^I$ is trivial. This
property becomes essential \eg\ when one wants to perform a
gauge-fixing 
of a part of $\k$-symmetry that transforms $\theta^I$. Then,
$Q_I$ has to be supplemented with a compensating gauge transformation,
which can not be allowed to interfere with cohomology. Consider an
infinitesimal ``deformation'' of the supersymmetry generator by a
$\k$-transformation, $\tilde Q_I=Q_I+M_I^\AA t_\AA$, where
$t_\AA$ are generators of some gauge transformations labelled by the
index $\AA$, and $M_I^\AA$ are infinitesimal
parameters. If 
$Q_I\Xi_\AA^{JK}={1\over2}\d_I^KQ_L\Xi_\AA^{JL}=\d_I^Kt_\AA\Sigma^J$ as
above, one can define
$\tilde\Sigma^I=\Sigma^I-M_J^\AA\Xi_\AA^{IJ}$, 
and still have $\tilde Q_I\tilde\Sigma^J=\delta_I^JS$. A finite
deformation, as when gauge-fixing is performed, will require the
discussion to be extended to an infinite sequence of descent
equations.

An interesting parallel to topological string theory can be observed
when one tries to construct a $\Sigma^I$ that is ``as $\k$-invariant
as possible'', order by order in fermions.
An Ansatz would, apart from the expression above, include terms that
are independent of $\theta$,
$$
\Sigma^I=\int d^3\xi\sqrt
g\theta^I+\int((C+k\Omega)\theta^I+R^I)\punkt\eqn
$$
Here, $R$ is a 3-form with $Q_IR=0$, and the the term containing
$\Omega$ modifies eq. (\ExactAction) with a purely topological term,
$$
Q_I\Sigma^J=\d^J_I(S+k\Omega)\punkt\Eqn\ExactActionMod
$$
Using elements of the calculation yielding $\k$-symmetry of the
action, one finds
$$
\d_\k\Sigma^I=\int i_\k(({\star}1+C+k\Omega)\w d\theta^I+dR^I)\punkt\eqn
$$
Invariance at lowest order can be achieved if $k=1$ and
$R^I_{abc}=-\sd_{abc\a}\psi^{\a I}$, in which case the lowest order
variation becomes $\int 2(\Pi_+\k^I)^0=0$, which is seen from the
decomposition (\GammaExpression) of the projection matrix in $G_2$ tensors. 
However, exact cancellation
to all orders
is not possible by addition of further terms in $R^I$. Again, of
course, the non-zero terms in the variation are trivial.
The relation (\ExactActionMod), with $k=1$, is the exact
correspondence to the fact that in topological string theory, the
BRST-trivial object is the action plus the integral of the K\"ahler
form, which is obtained from $\Omega$ on dimensional reduction.

\section\TopologicalMTheory{Topological Membranes in Topological
M-theory}We have shown how a supermembrane in seven dimensions with
euclidean signature can be turned into a topological theory.
It would be interesting to study the quantum mechanical properties of
the topological membrane theory, and investigate to what extent the
quantum theory reproduces topological M-theory. The best framework for
doing this would be one including a proper set of auxiliary fields
that makes the symmetries of the theory valid of-shell. It seems much
harder to reach such a formulation in the present situation than for
the usual world-sheet supersymmetric sigma model on which topological
string theory is based. 

It is clear that associative cycles [\Joyce] are solutions of the theory. These
are calibrating cycles for the 3-form $\Omega$. An easy way to see
that associative cycles are supersymmetric is to partially fix gauge
for $\k$-symmetry by demanding $\theta^I=0$. The supersymmetry,
including a compensating
$\k$-transformation, on the remaining fermions becomes
$$
\d_\e\psi^{\a I}=-{z^\a\over1-y}\e^I\komma\Eqn\SevenSupersymmetry
$$
where $y={1\over6\sqrt g}\e^{ijk}\s_{ijk}$, 
$z_\a={1\over6\sqrt g}\e^{ijk}\sd_{ijk\a}$.
A configuration is supersymmetric if $z^\a=0$, giving the
possibilities $y=\pm1$, and if eq. (\SevenSupersymmetry) is to be well
behaved only $y=-1$ is possible. With a non-zero Wess--Zumino term in the
membrane action we are actually dealing with a generalised calibration,
see \eg\ refs. [\TownsendCalM,\GutowskiPapadopAdSCal]. It is however
of a trivial type since 
the bosonic 3-form is closed and hence the Wess--Zumino term 
contributes equally to all cycles
minimal or not. 

Looking for local observables seems more problematic. In the A-model,
considering collapsed, point-like, world-sheets is straightforward,
and cohomology of the BRST-operator is directly translated into
cohomology for a de Rahm-complex for the Calabi--Yau manifold. In the
present situation, we have to take $\k$-symmetry into account, with
its projection that depends on the orientation of the embedded
world-volume. We have not yet been able to address this question in a
constructive way, and thus can not present a direct connection between
observables for the topological membrane and Hitchin's theory.

It is clear that a
double dimensional reduction of the topological membrane produces the
strings of the topological A-model, although formulated in a space
supersymmetric rather than world-sheet supersymmetric
way. Associative cycles will map to holomorphic cycles. For the same
reasons as above, we are not able to make a corresponding statement
concerning local observables (although investigating this question for
the A-model starting from a Green--Schwarz formulation 
might give some insight)\foot\dagger{Such a formulation will be
possible directly in six dimensions for both the A- and B-models. One
has a priori an SL(2) doublet of complex supersymmetries, of which
different real combinations may be chosen.}.
A direct reduction will of course
give A-model 2-branes. These are not D-branes. An A-model D2-brane
must be represented by a 3-brane in $D=7$, since the boundary of an
open membrane winding the compactified circle also winds. This, along
with the existence of the dual form ${\star}\Omega$, makes it clear
that 3-branes, on which the membranes may end, are needed in
topological M-theory. The 3-branes, living on the same superspace,
should support a world-volume 2-form potential, with a 3-form field
strength. This field, that can be dualised to a scalar, accounts for
the correct matching of bosonic and fermionic degrees of freedom.
An interesting observation, on which we would like to elaborate in the
future, is that although $(\g_{abc})_{\hat\a\hat\b}$ is symmetric in
spinor indices, and thus cannot be used in a dimension-0 component of 
the 5-form field strength
for the 4-form potential coupling to the 3-brane, there exists a
closed 5-form constructed from $G_2$-invariant tensors.

It would be a great step forward to find a good set of auxiliary
fields for the membrane theory, that would allow for an off-shell
formulation, and hopefully make quantisation more manageable. Although
this, in general backgrounds, would probably be to ask too much, it is
maybe not unrealistic to hope that the $G_2$ structure would help.
It turns spinors into scalars and vectors, and even $\k$-symmetry can
be parametrised covariantly, as in eq. (\CovariantKappa). One possible
starting point could be the construction of a super-diffeomorphism
algebra on the 
world-volume containing an $SL(2)$ doublet of fermionic vector generators,
similar to what one obtains after gauge-fixing the scalar part of
$\k$-symmetry.

Although we do not claim to have a microscopic definition of
topological M-theory, we hope that the present work represents a step
in that direction. Maybe it can be a point of departure for a refined
formulation, where urgent questions, such as the connection to
Hitchin's theory of $G_2$ moduli, can be answered. Such a formulation
might also give valuable insight into the question of how membrane
functional integrals are performed (see \eg\ the discussion in
ref. [\HarveyMoore]). Earlier experience of instanton counting on
compact submanifolds have shown that naive counting of membrane
configurations may lead to incorrect results 
[\PiolineNicolaiPlefkaWaldron,\SuginoVanhove], and a proper theory of
topological membranes may be a place where such issues can be
addressed in a precise manner.

\vfill\eject

\appendix{Some details on $G_2$ tensors}We use \eg\ the 
expressions $\s_{a,a+1,a+3}=1$ (where indices are counted modulo 7), giving
$\sd_{a,a+1,a+2,a+5}=1$. $\sd$ is the octonionic associator, 
$[e_a,e_b,e_c]=(e_ae_b)e_c-e_a(e_be_c)=-2\sd_{abcd}e_d$.
Useful relations between octonionic structure constants:
$$
\eqalign{
\s_{acd}\s_b{}^{cd}&=6\d_{ab}\komma\cr
\sd_{abde}\s_c{}^{de}&=-4\s_{ijk}\komma\cr
\s_{abc}\s^{abc}&=42
                                       \komma\cr
\s_{abe}\s_{cd}{}^e&=2\d^{ab}_{cd}-\sd_{abcd}\komma\cr
\s_{abf}\sd_{cde}{}^f&=6\d^{[a}_{[c}\s^{b]}{}_{de]}\komma\cr
\sd_{abcg}\sd_{def}{}^g&=6\d^{abc}_{def}-3\d^{[a}_{[d}\sd^{bc]}{}_{ef]}
          -3\s_{[ab}{}^{[d}\s_{c]}{}^{ef]}\komma\cr
\sd_{abef}\sd_{cd}{}^{ef}&=8\d^{ab}_{cd}-2\sd_{abcd}\punkt\cr
}\eqn
$$
The last of these relations can be used to find projections on the 7- and
14-dimensional vector spaces in $21\rightarrow14\oplus7$ under 
$Spin(7)\rightarrow G_2$ as
$$
\eqalign{
\Pi^{(14)}{}_{ab}{}^{cd}&=\Fr23(\d_{ab}^{cd}+\fr4\sd_{ab}{}^{cd})\komma\cr
\Pi^{(7)}{}_{ab}{}^{cd}&=\fr3(\d_{ab}^{cd}-\fr2\sd_{ab}{}^{cd})\punkt\cr
}\eqn
$$
It can be noted that
$\s_{abc}$, seen as a set of seven matrices $(\s_a)_{bc}$, are in the
7-dimensional subspace: $(\Pi^{(14)}\s_a)_{bc}=0$, and actually
provide a basis for it.

Consider the split of the octonions $\OO$ as $\HH\oplus\HH$, and write 
$x=\xi+j\eta$, where $\xi$ and $\eta$ are quaternions and $j$ is an
imaginary unit orthogonal to $\HH$. The octonionic
multiplication is encoded in terms of the quaternions by the
multiplication rules $ja=a^*j$, $(ja)b=j(ba)$ for all $a,b\in\HH$.
Then 
$$
xx'=\xi\xi'-\eta'\eta^*+j(\xi^*\eta'+\xi'\eta)\punkt\Eqn\OintermsofH
$$ 
We want to
examine which of the rotations in $SO(3)\times SO(4)$ acting on
imaginary octonions and preserving this split are automorphisms, \ie,
belong to $G_2$. The rotations are parametrised as
$\xi\rightarrow\s^*\xi\s$, $\eta\rightarrow e^*\eta e'$, where all
three parameters are unit quaternions. A direct check with
eq. (\OintermsofH)
yields that the necessary condition for this to be an automorphism 
is $\s=e$, verifying that the common subgroup of this $SO(3)\times
SO(4)$ and $G_2$ is $SU(2)\times SU(2)$, and that the twisting---the
identification of world-volume $SO(3)$ rotations with a transverse
$SU(2)$---takes place. 

The remaining part of the $G_2$ algebra transforms as $(4,2)$, and is realised
infinitesimally with a ``vector-spinor'' $h_i$, $i=1,2,3$, 
in $\HH^{\otimes3}$ with
$e_ih_i=0$. The transformations are $\d\xi=e_i\eta h_i^*$,
$\d\eta=e_i\xi h_i\,(=-2\xi_ih_i)$, 
and the derivation property may be checked explicitly.

The split into two quaternions can also be seen
as a split in four complex numbers with imaginary unit $j$. With
$x=z_0+z^ie_i$, the multiplication table is
$xx'=z_0z'_0-z^i\bar z'_i
+(z_0z'^i+z^i\bar z'_0+\e^{ijk}\bar z_j\bar z'_k)e_i$, in which
$SU(3)\subset G_2$ is a manifest automorphism. 
The rest of the automorphisms are parametrised by $\lambda^i$,
$\bar\lambda_i$ in $3\oplus\bar3$, acting as 
$\d z_0=\lambda^i\bar z_i-\bar\lambda_iz^i$, 
$\d z^i=\lambda^i(z_0-\bar z_0)+\e^{ijk}\bar\lambda_j\bar z_k$.

\appendix{3-forms in superspace}The field strength $G$ is related to
the potential $C$ in the conventional way
$$
G = dC \hskip.7cm \Rightarrow \hskip.7cm G_{ABCD} =
4\delta_{[A}C_{BCD)} + 6T_{[AB}^{\phantom{[AB}F}C_{|F|CD)}\komma \Eqn\potC
$$
where the indices in capital letters are the entire superspace
indices. The bracket $[*)$ denotes a weighted symmetrisation, i.e.,
anti-symmetrisation or symmetrisation depending on if one considers
bosonic or the fermionic form indices. Using the fact that in
a flat background, the only non-vanishing components of $G_{ABCD}$ and
$T_{AB}^{\phantom{AB}C}$ are $G_{ab,\hat{\gamma}I,\hat{\delta}J} =
-2\varepsilon_{IJ}(\gamma_{ab})_{\hat{\gamma}\hat{\delta}}$ and
$T_{\hat{\alpha}I,\hat{\beta}J}^{\phantom{\hat{\alpha}I,\hat{\beta}J}c}
= 2\varepsilon_{IJ}(\gamma^c)_{\hat{\alpha}\hat{\beta}}$,
respectively, the equation for $C_{ABC}$ can be solved. The solution
we are interested in has the property that the only coordinate
dependence is through the seven-dimensional fermionic coordinates
$\psi^{\alpha I}$. By looking at the group representation structures
of the different components of $C$, we made an Ansatz for the
potential, where the Ansatz parameters were fixed by Eq. \potC. Due to
the invariance under the gauge transformation $\delta C = d\Lambda$,
some of the Ansatz parameters are free, which we for simplicity set to
zero. Since $so(7)$ is no longer a valid symmetry under some of the
gauge transformations, we have made the Ansatz using $G_2$ invariants
and $\psi^{\alpha I}$ as ingredients, which means that we use a flat
space or work to lowest order in curvatures.
The potential we have found can
be written as $C^{(k)} = C + k\Omega$, where
$$
\eqalign{
C_{abc} & = 0 \cr
C_{ab,0I} & = \e_{IL} \psi^{\delta L} 2\sigma_{ab\delta} \cr
C_{ab,\alpha I} & = \e_{IL} \psi^{\delta L}
(2\delta_{\alpha\delta}^{ab} + \star\sigma_{ab\alpha\delta}) \cr
C_{a,0I,0J} & = \e_{I(L}\e_{|J|M)} \psi^{\delta L}
\psi^{\epsilon M} (-4)\sigma_{a\delta\epsilon} \cr
C_{a,\alpha I,0J} & =  \e_{I(L}\e_{|J|M)} \psi^{\delta L}
\psi^{\epsilon M} (-4\delta^{a\alpha}_{\delta\epsilon} -
\Fr{2}{3}\star\sigma_{a\alpha\delta\epsilon}) \cr
C_{a,\alpha I,\beta J} & = \e_{I(L}\e_{|J|M)} \psi^{\delta L}
\psi^{\epsilon M} (-\Fr{4}{3}\sigma_{a\delta\epsilon}\delta_{\alpha\beta} -
\Fr{8}{3}\sigma_{a\alpha\delta}\delta_{\beta\epsilon}) \cr
C_{0I,0J,0K} & = \e_{I(L}\e_{|J|M}\e_{|K|N)} \psi^{\delta L}
\psi^{\epsilon M} \psi^{\phi N} 8\sigma_{\delta\epsilon\phi} \cr
C_{\alpha I,0J,0K} & = \e_{IL}\e_{J(M}\e_{|K|N)} \psi^{\delta L}
\psi^{\epsilon M} \psi^{\phi N} (-\Fr{8}{3})
\star\sigma_{\alpha\delta\epsilon\phi} \cr
C_{\alpha I,\beta J,0K} & = \e_{IL}\e_{JM}\e_{KN} \psi^{\delta L}
\psi^{\epsilon M} \psi^{\phi N} (\Fr{8}{3}\delta_{\alpha\beta}
\sigma_{\delta\epsilon\phi} + 8\sigma_{\alpha\epsilon\phi}
\delta_{\beta\delta}) \cr
C_{\alpha I,\beta J,\gamma K} & = \e_{IL}\e_{JM}\e_{KN}
\psi^{\delta L} \psi^{\epsilon M} \psi^{\phi N} 8\delta_{\alpha\phi}
\star\sigma_{\beta\gamma\delta\epsilon} \cr
}\Eqn\CIs
$$
$$
\eqalign{
\Omega_{abc} & = \sigma_{abc} \cr
\Omega_{ab,0I} & = \e_{IL} \psi^{\delta L} (-2)\sigma_{ab\delta} \cr
\Omega_{ab,\alpha I} & = \e_{IL} \psi^{\delta L}
(-2\delta_{\alpha\delta}^{ab} + \star\sigma_{ab\alpha\delta}) \cr
\Omega_{a,0I,0J} & = \e_{I(L}\e_{|J|M)} \psi^{\delta L}
\psi^{\epsilon M} 4\sigma_{a\delta\epsilon} \cr
\Omega_{a,\alpha I,0J} & =  \e_{I(L}\e_{|J|M)} \psi^{\delta L}
\psi^{\epsilon M} (4\delta^{a\alpha}_{\delta\epsilon} +
2\star\sigma_{a\alpha\delta\epsilon}) \cr
\Omega_{a,\alpha I,\beta J} & = \e_{I(L}\e_{|J|M)} \psi^{\delta L}
\psi^{\epsilon M} (\Fr{8}{3}\delta_{a\alpha}\sigma_{\beta\delta\epsilon}
- \Fr{4}{3}\sigma_{a\delta\epsilon}\delta_{\alpha\beta} -
\Fr{8}{3}\sigma_{a\alpha\delta}\delta_{\beta\epsilon}) \cr
\Omega_{0I,0J,0K} & =  \e_{I(L}\e_{|J|M}\e_{|K|N)} \psi^{\delta L}
\psi^{\epsilon M} \psi^{\phi N} (-8)\sigma_{\delta\epsilon\phi} \cr
\Omega_{\alpha I,0J,0K} & = \e_{IL}\e_{J(M}\e_{|K|N)} \psi^{\delta L}
\psi^{\epsilon M} \psi^{\phi N} 4 \star\sigma_{\alpha\delta\epsilon\phi}
\cr
\Omega_{\alpha I,\beta J,0K} & =  \e_{IL}\e_{JM}\e_{KN} \psi^{\delta L}
\psi^{\epsilon M} \psi^{\phi N} (-\Fr{8}{3}\delta_{\alpha\beta}
\sigma_{\delta\epsilon\phi} - \Fr{56}{3}\sigma_{\alpha\epsilon\phi}
\delta_{\beta\delta}) \cr
\Omega_{\alpha I,\beta J,\gamma K} & = \e_{IL}\e_{JM}\e_{KN}
\psi^{\delta L} \psi^{\epsilon M} \psi^{\phi N} 8\delta_{\alpha\phi}
\star\sigma_{\beta\gamma\delta\epsilon} \cr
}\Eqn\OmegaIs
$$

\noindent
and $k$ is a free parameter. Symmetrisation in composite fermionic
indices is implicitly understood in eqs. (\CIs) and
(\OmegaIs). Eq. (\OmegaIs) can of course be obtained directly (modulo
an exact form) by expanding the bosonic differentials in 
$\Omega={1\over6}dx^p\w dx^n\w dx^me_p{}^ce_n{}^be_m{}^a\s_{abc}$ using the
vielbeins of eq. (\FormOfVielbeins).

The fact that $\theta^I$ are $G_2$-invariant should make it clear that
the proof of local $\theta$-independence of the lagrangian, on which the
BRST-exactness relies, may be generalised to curved backgrounds,
involving modifications of the explicit forms of the supervielbeins of
eq. (\FormOfVielbeins) and the super-3-forms of eqs. (\CIs) and (\OmegaIs).
\vfill\eject

\acknowledgements
MC wants to thank Paul Howe and Per Sundell for discussions and
clarifying comments. The research project is financed in part by the
Swedish Science Council and by EU contract 
MRTN-CT-2004-512194.

\refout

\end